\documentclass[prl,showpacs,twocolumn]{revtex4}
\usepackage{amssymb}
\usepackage{amsmath}
\usepackage{graphicx}

\begin{document}

\title{General Canonical Quantum Gravity Theory and that of the Universe and
General Black Hole}
\author{C. Huang $^{\!\!1,2}$}
\email{c.huang0@hotmail.com}
\author{Yong-Chang Huang $^{\!\!3}$}
\email{ychuang@bjut.edu.cn}
\author{Xinfei Li$^{3,}$}
\email{xfli@bjut.edu.cn}
\affiliation{$^1$Lawrence Berkeley National Laboratory, 1 Cyclotron Road, Berkeley CA
94720, USA\\
$^2$Department of Physics and Astronomy, Purdue University, 525 Northwestern
Avenue, W.Lafayette, IN 47907-2036, USA\\
$^3$Institute of Theoretical Physics, Beijing University of Technology,
Beijing 100124, China\\
}
\date{\today }

\begin{abstract}
This paper gives both a general canonical quantum gravity theory and the
general canonical quantum gravity theories of the Universe and general black
hole, and discovers the relations reflecting symmetric properties of the
standard nonlinear gravitational Lagrangian, which are not relevant to any
concrete metric models. This paper concretely shows the general commutation
relations of the general gravitational field operators and their zeroth,
first, second and third style, respectively, of high order canonical
momentum operators for the general nonlinear system of the standard
gravitational Lagrangian, and then has finished all the four styles of the
canonical quantization of the standard gravity.

\end{abstract}

\maketitle

\section{I. Introduction}

Quantum gravity is a domain of theoretical physics for exploring to describe
gravity on the basis of principles of quantum mechanics \cite{Rove1}, when
near compact astrophysical objects, there are the strong effects of gravity.

The current gravitational theory is based on Einsteinian general relativity
in classical physics, and the other three physical fundamental forces are
represented in quantum field theory, they are very different formalisms for
explaining different physical phenomena \cite{Grif}. Quantum gravity is
necessary when people's studies are from classical physics to quantum
physics \cite{Wald,Feyn,gary,hbb}.

Quantum gravity may reconcile general relativity with quantum mechanics,
there exist difficulties when quantum field theory is used to gravitational
theory by graviton bosons \cite{Zee}, and the deduced theory is not
renormalizable (e.g., the theory shows infinite values of observable
quantities, for instance, the masses of particles). Therefore, theorists
have done a lot of research works in order to overcome the problems of
quantum gravity. Superstring theory unifying gravity with the other three
fundamental forces and loop quantum gravity no such attempt are the good
candidates for overcoming the problems of quantum gravity \cite{Penr}, and
both superstring theory and loop quantum gravity all quantize the
gravitational fields.

Quantum gravity shows the quantum behavior of the gravitational field, and
superstring theory of unifying grand unified theory and gravitational theory
may be viewed as a theory of everything. The investigations of quantum
gravity are domains having different approaches for the unification.

Up to now, no less, at least, than 16 major interesting approaches for
quantum gravity have been shown in the literature \cite{Bryce} in
alphabetical order as follows:

Affine quantum gravity \cite{Klau}; Asymptotic quantization \cite{Gero,Asht}%
; Canonical quantum gravity \cite{DeWi, Isha, Ash, Espo}; Condensed-matter
view \cite{Volo}; Manifestly covariant quantization \cite{Misn, DeW, Hoof,
Goro, Avra, Vilk, Canf}; Euclidean quantum gravity \cite{Gibb, Haw}; Lattice
formulation \cite{Meno, Carf1}; Loop space representation \cite{Rove, Rov};
Non-commutative geometry \cite{Grac}; Quantum topology \cite{Isha1}, \cite%
{Whee}; Renormalization group and asymptotic safety \cite{Reut, Laus};
R-squared gravity \cite{Stel}; String and brane theory \cite{Witte, Horo,
Barv,haa}; Supergravity \cite{Free, Nieu}; Triangulations \cite{Gion, Ambj,
Loll} and null-strut calculus \cite{Khey}; Twistor theory \cite{Penro,
Penro1}.

Quantum gravitational effects evidently show at scales near the Planck scale
and equivalently far larger energy, which is far larger than that of current
high energy particle accelerators. Consequently, there are not experimental
data distinguishing the proposed competing theories, but thought
experimental methods are presented as the testing methods of the competing
theories \cite{Bose, Marl, Nemi}.

The quantization of gravity, up to now, remains a formidable problem for
physicists. Although superstring theory has made some progress in quantizing
gravity, many profound questions still remain unanswered \cite{bbs}.

Meshing all these theories at all energy scales is relevant to the different
assumptions how the universe works. General relativity shows that "spacetime
tells matter how to move; matter tells spacetime how to curve."\cite{Wheel}.
Quantum field theory is formulated according to special relativity in the
flat spacetime. When treating gravitation as a simple quantum field, which
will result in that the theory is not renormalizable \cite{Zee}. Quantizing
gravity becomes key challenges, and is no longer applicable in flat
spacetime \cite{Wald1}.

Quantum gravitational theory is widely hoped to understand the origin of the
universe and the behaviors of black holes \cite{Hawk1}. Some different
quantum physics systems are investigated \cite{abcd}, and Minimal area
surfaces dual to Wilson loops are studied \cite{Huang1}.

The appearance of singularity of infinite large of spacetime curvature in
general relativity (meaning its structure has a microscopic scale) requires
the establishment of a complete theory of quantum gravity. The quantum
gravitational theory needs to be able to describe the conditions inside
black holes and in the very early universe, where gravity and the related
spacetime geometry need to be described in quantized formulism. Despite lots
of efforts by physicists and the developments of some potential candidate
theories, humans have yet to come up with a complete and self-consistent
theory of quantum gravity. This paper wants to solve the problems in order
to give a complete and self-consistent general theory of quantum gravity.

Especially, after having done almost all sorts of the great efforts, e.g.,
see refs. \cite{acb1,acb2,acb3,acb4}, theoretical physicists, all over the
world, doing research on modern field theory still don't know and haven't
found the general theory of quantum gravity. All these very hard problems
can be called as quantum gravity puzzles. In order to solve the puzzles of
quantum gravity, so the experts in quantum gravity try to study different
kinds of gravity, specially, the gravity of black holes and their related
quantization, try to find out the most general characteristics of quantum
gravity, so as to establish the general theory of quantum gravity.

In this paper, according to the general method of canonical field theory in
modern quantum field theory, we not only find and establish the general
canonical quantum gravity theory, but also give their application to the
Universe and general black hole, i.e., we deduce the geneal canonical
quantum gravity theories of the Universe and general black hole.

The arrangement of this paper is: Sect. 2 gives canonical conjugate momentum
opretors corresponding gravitational field operators, Sect. 3 studies energy
of gravitational field of the standard gravtational Lagrangian, Sect.4 shows
further investigations of quantization of quantum gravitational fields,
Sect.5 investigates commutation relation for gravitational fields and the
first style of momenta, Sect. 6 shows commutation relation for gravitational
fields and the second style of momenta, Sect. 7 studies commutation relation
for graviational fields and the third style of momenta, Sect.8 shows the
general canonical quantum gravity of the Universe, and Sect. 9 gives the
general canonical quantum gravity of general black hole, Sect. 10 is summary
and conclusion.

\section{II. Canonical conjugate momentum opretors corresponding
gravitational field operators}

We generally consider granvitational field operators

\begin{equation}
\hat{g}_{\mu \nu }(x)=\hat{g}_{\nu \mu }(x),\mu ,\nu =0,1,2,3,  \tag{2.1}
\end{equation}%
thus there are ten independent components. \ \

We need to specially stress that our investigations in this paper are very
general for the standard gravitational field Lagrangian that has gotten the
huge successes in classical physics, and are not dependent on any concrete
metric model through the whole paper. Thus, the studies of this paper is of
the general theory of quantum gravity.

When we generalize coordinate operators and canonical momentum operators of
finite degrees of freedom to gravity field operators and its canonical
conjugate momenta of infinite degrees of freedom, physics consistence
requires us to generalize commutation relation of coordinate operators and
momentum operators in the first quantization to commutation relation of
gravity field operators and its canonical conjugate momenta

\begin{equation}
\lbrack \hat{g}_{\mu \nu }(\mathbf{x},t),\hat{\pi}^{\alpha \beta }(\mathbf{x}%
^{\prime },t^{\prime })]_{t=t^{\prime }}=i\Delta _{\mu \nu }^{\alpha \beta
}\delta (\mathbf{x-x}^{\prime }),  \tag{2.2}
\end{equation}%
where $\Delta _{\mu \nu }^{\alpha \beta }$ is a general operator function
decided by satisfying some conditions of this system, the corresponding
canonical momentum is

\begin{equation}
\pi ^{\mu \nu }(\mathbf{x},t)=\frac{\partial \mathcal{L}}{\partial \partial
_{t}g_{\mu \nu }(\mathbf{x},t)}\rightarrow \hat{\pi}^{\mu \nu }(\mathbf{x}%
,t)=\frac{\partial \hat{\mathcal{L}}}{\partial \partial _{t}\hat{g}_{\mu \nu
}(\mathbf{x},t)},  \tag{2.3}
\end{equation}%
\begin{equation}
\mathcal{L}=\kappa R=\kappa g^{\alpha \beta }R_{\alpha \beta },  \tag{2.4}
\end{equation}%
where $\mathcal{L}$, $R$ and $\kappa $ are the standard Lagrangian density
of general relativity \cite{Jerzy}, the scalar curvature and the coupling
constant of the general gravity system, respectively.

Generally taking the standard Lagrangian density (2.4) of this system and
using eq.(2.3), we get

\begin{equation}
\pi ^{\mu \nu }(\mathbf{x},t)=\frac{\partial \mathcal{L}}{\partial \partial
_{t}g_{\mu \nu }(\mathbf{x},t)}=\kappa g^{\alpha \beta }\frac{\partial
R_{\alpha \beta }}{\partial \partial _{t}g_{\mu \nu }(\mathbf{x},t)}.
\tag{2.5}
\end{equation}

Putting

\begin{equation}
R_{\alpha \beta }=\Gamma _{\alpha \sigma ,\beta }^{\sigma }-\Gamma _{\alpha
\beta ,\sigma }^{\sigma }+\Gamma _{\rho \beta }^{\sigma }\Gamma _{\alpha
\sigma }^{\rho }-\Gamma _{\rho \sigma }^{\sigma }\Gamma _{\alpha \beta
}^{\rho }  \tag{2.6}
\end{equation}%
into eq.(2.5), it follows that

\begin{equation}
\pi ^{\mu \nu }(\mathbf{x},t)=\kappa g^{\alpha \beta }\frac{\partial (\Gamma
_{\alpha \sigma ,\beta }^{\sigma }-\Gamma _{\alpha \beta ,\sigma }^{\sigma
}+\Gamma _{\rho \beta }^{\sigma }\Gamma _{\alpha \sigma }^{\rho }-\Gamma
_{\rho \sigma }^{\sigma }\Gamma _{\alpha \beta }^{\rho })}{\partial \partial
_{t}g_{\mu \nu }(\mathbf{x},t)}.  \tag{2.7}
\end{equation}

Using connection

\begin{equation}
\Gamma _{\rho \beta }^{\sigma }=\frac{1}{2}g^{\sigma \gamma }(g_{\gamma
\beta ,\rho }+g_{\rho \gamma ,\beta }-g_{\rho \beta ,\gamma }),  \tag{2.8}
\end{equation}%
eq.(2.7) can be rewritten as

\begin{equation*}
\pi ^{t\mu \nu }(\mathbf{x},t)=\kappa g^{\alpha \beta }\frac{\partial }{%
\partial \partial _{t}g_{\mu \nu }(\mathbf{x},t)}[
\end{equation*}

\begin{eqnarray*}
&&\frac{1}{2}g_{,\beta }^{\sigma \gamma }(g_{\gamma \sigma ,\alpha
}+g_{\alpha \gamma ,\sigma }-g_{\alpha \sigma ,\gamma })-\frac{1}{2}%
g_{,\sigma }^{\sigma \gamma }(g_{\gamma \beta ,\alpha }+g_{\alpha \gamma
,\beta }-g_{\alpha \beta ,\gamma }) \\
&&+\frac{1}{2}g^{\sigma \gamma }(g_{\gamma \beta ,\rho }+g_{\rho \gamma
,\beta }-g_{\rho \beta ,\gamma })\frac{1}{2}g^{\rho \gamma }(g_{\gamma
\sigma ,\alpha }+g_{\alpha \gamma ,\sigma }-g_{\alpha \sigma ,\gamma })
\end{eqnarray*}

\begin{equation}
-\frac{1}{2}g^{\sigma \gamma }(g_{\gamma \sigma ,\rho }+g_{\rho \gamma
,\sigma }-g_{\rho \sigma ,\gamma })\frac{1}{2}g^{\rho \gamma }(g_{\gamma
\beta ,\alpha }+g_{\alpha \gamma ,\beta }-g_{\alpha \beta ,\gamma })].
\tag{2.9}
\end{equation}

Using formulae

\begin{equation}
g^{\sigma \mu }g_{\mu \beta ,\rho }=-g_{,\rho }^{\sigma \mu }g_{\mu \beta
},g^{\sigma \mu }g_{\mu \beta ,\rho }g^{\beta \alpha }=-g_{,\rho }^{\sigma
\mu }g_{\mu \beta }g^{\beta \alpha }=-g_{,\rho }^{\sigma \alpha },
\tag{2.10}
\end{equation}

eq.(2.9) can be reexpressed as

\begin{equation*}
\pi ^{t\mu \nu }(\mathbf{x},t)=\kappa g^{\alpha \beta }[\frac{-1}{2}%
g^{\sigma \tau }\delta _{\beta }^{t}\delta _{\tau }^{\mu }\delta
_{\varepsilon }^{\nu }g^{\varepsilon \gamma }(g_{\gamma \sigma ,\alpha
}+g_{\alpha \gamma ,\sigma }-g_{\alpha \sigma ,\gamma })
\end{equation*}

\begin{eqnarray*}
&&-\frac{1}{2}g^{\sigma \tau }g_{\tau \varepsilon ,\beta }g^{\varepsilon
\gamma }(\delta _{\alpha }^{t}\delta _{\gamma }^{\mu }\delta _{\sigma }^{\nu
}+\delta _{\sigma }^{t}\delta _{\alpha }^{\mu }\delta _{\gamma }^{\nu
}-\delta _{\gamma }^{t}\delta _{\alpha }^{\mu }\delta _{\sigma }^{\nu }) \\
&&+\frac{1}{2}g^{\sigma \tau }\delta _{\sigma }^{t}\delta _{\tau }^{\mu
}\delta _{\varepsilon }^{\nu }g^{\varepsilon \gamma }(g_{\gamma \beta
,\alpha }+g_{\alpha \gamma ,\beta }-g_{\alpha \beta ,\gamma }) \\
&&+\frac{1}{2}g^{\sigma \tau }g_{\tau \varepsilon ,\sigma }g^{\varepsilon
\gamma }(\delta _{\alpha }^{t}\delta _{\gamma }^{\mu }\delta _{\beta }^{\nu
}+\delta _{\beta }^{t}\delta _{\alpha }^{\mu }\delta _{\gamma }^{\nu
}-\delta _{\gamma }^{t}\delta _{\alpha }^{\mu }\delta _{\beta }^{\nu })+%
\frac{1}{4}g^{\sigma \gamma }(\delta _{\rho }^{t}\delta _{\gamma }^{\mu
}\delta _{\beta }^{\nu } \\
&&+\delta _{\beta }^{t}\delta _{\rho }^{\mu }\delta _{\gamma }^{\nu }-\delta
_{\gamma }^{t}\delta _{\rho }^{\mu }\delta _{\beta }^{\nu })g^{\rho \gamma
^{\prime }}(g_{\gamma ^{\prime }\sigma ,\alpha }+g_{\alpha \gamma ^{\prime
},\sigma }-g_{\alpha \sigma ,\gamma ^{\prime }})+\frac{1}{4}g^{\sigma \gamma
}( \\
&&g_{\gamma \beta ,\rho }+g_{\rho \gamma ,\beta }-g_{\rho \beta ,\gamma
})g^{\rho \gamma ^{\prime }}(\delta _{\alpha }^{t}\delta _{\gamma ^{\prime
}}^{\mu }\delta _{\sigma }^{\nu }+\delta _{\sigma }^{t}\delta _{\alpha
}^{\mu }\delta _{\gamma ^{\prime }}^{\nu }-\delta _{\gamma ^{\prime
}}^{t}\delta _{\alpha }^{\mu }\delta _{\sigma }^{\nu })-
\end{eqnarray*}%
\begin{equation*}
\frac{1}{4}g^{\sigma \gamma }(\delta _{\rho }^{t}\delta _{\gamma }^{\mu
}\delta _{\sigma }^{\nu }+\delta _{\sigma }^{t}\delta _{\rho }^{\mu }\delta
_{\gamma }^{\nu }-\delta _{\gamma }^{t}\delta _{\rho }^{\mu }\delta _{\sigma
}^{\nu })g^{\rho \gamma ^{\prime }}(g_{\gamma ^{\prime }\beta ,\alpha
}+g_{\alpha \gamma ^{\prime },\beta }-g_{\alpha \beta ,\gamma ^{\prime }})
\end{equation*}

\begin{equation}
-\frac{1}{4}g^{\sigma \gamma }(g_{\gamma \sigma ,\rho }+g_{\rho \gamma
,\sigma }-g_{\rho \sigma ,\gamma })g^{\rho \gamma ^{\prime }}(\delta
_{\alpha }^{t}\delta _{\gamma ^{\prime }}^{\mu }\delta _{\beta }^{\nu
}+\delta _{\beta }^{t}\delta _{\alpha }^{\mu }\delta _{\gamma ^{\prime
}}^{\nu }-\delta _{\gamma ^{\prime }}^{t}\delta _{\alpha }^{\mu }\delta
_{\beta }^{\nu })].  \tag{2.11}
\end{equation}

For convenience and simplicity and no losing generality, we transform
eq.(2.11) as coveriant 3-order tensor

\begin{equation*}
\pi _{\lambda \theta \chi }(\mathbf{x},t)=g_{\tau \lambda }g_{\mu \theta
}g_{\nu \chi }\pi ^{\tau \mu \nu }(\mathbf{x},t)=
\end{equation*}%
\begin{equation*}
\kappa \lbrack \frac{-1}{2}g_{\tau \lambda }g_{\mu \theta }g_{\nu \chi
}g^{\alpha \beta }g^{\sigma \tau ^{\prime }}\delta _{\beta }^{\tau }\delta
_{\tau ^{\prime }}^{\mu }\delta _{\varepsilon }^{\nu }g^{\varepsilon \gamma
}(g_{\gamma \sigma ,\alpha }+g_{\alpha \gamma ,\sigma }-g_{\alpha \sigma
,\gamma })
\end{equation*}

\begin{eqnarray*}
&&-\frac{1}{2}g_{\tau \lambda }g_{\mu \theta }g_{\nu \chi }g^{\alpha \beta
}g^{\sigma \tau ^{\prime }}g_{\tau ^{\prime }\varepsilon ,\beta
}g^{\varepsilon \gamma }(\delta _{\alpha }^{\tau }\delta _{\gamma }^{\mu
}\delta _{\sigma }^{\nu }+\delta _{\sigma }^{\tau }\delta _{\alpha }^{\mu
}\delta _{\gamma }^{\nu }-\delta _{\gamma }^{\tau }\delta _{\alpha }^{\mu
}\delta _{\sigma }^{\nu }) \\
&&+\frac{1}{2}g_{\tau \lambda }g_{\mu \theta }g_{\nu \chi }g^{\alpha \beta
}g^{\sigma \tau ^{\prime }}\delta _{\sigma }^{\tau }\delta _{\tau ^{\prime
}}^{\mu }\delta _{\varepsilon }^{\nu }g^{\varepsilon \gamma }(g_{\gamma
\beta ,\alpha }+g_{\alpha \gamma ,\beta }-g_{\alpha \beta ,\gamma }) \\
&&+\frac{1}{2}g_{\tau \lambda }g_{\mu \theta }g_{\nu \chi }g^{\alpha \beta
}g^{\sigma \tau ^{\prime }}g_{\tau ^{\prime }\varepsilon ,\sigma
}g^{\varepsilon \gamma }(\delta _{\alpha }^{\tau }\delta _{\gamma }^{\mu
}\delta _{\beta }^{\nu }+\delta _{\beta }^{\tau }\delta _{\alpha }^{\mu
}\delta _{\gamma }^{\nu }-\delta _{\gamma }^{\tau }\delta _{\alpha }^{\mu
}\delta _{\beta }^{\nu })
\end{eqnarray*}

\begin{eqnarray*}
&&+\frac{1}{4}g_{\tau \lambda }g_{\mu \theta }g_{\nu \chi }g^{\alpha \beta
}g^{\sigma \gamma }g^{\rho \gamma ^{\prime }}(\delta _{\rho }^{\tau }\delta
_{\gamma }^{\mu }\delta _{\beta }^{\nu }+\delta _{\beta }^{\tau }\delta
_{\rho }^{\mu }\delta _{\gamma }^{\nu }-\delta _{\gamma }^{\tau }\delta
_{\rho }^{\mu }\delta _{\beta }^{\nu })( \\
&&g_{\gamma ^{\prime }\sigma ,\alpha }+g_{\alpha \gamma ^{\prime },\sigma
}-g_{\alpha \sigma ,\gamma ^{\prime }})+\frac{1}{4}(g_{\gamma \beta ,\rho
}+g_{\rho \gamma ,\beta }-g_{\rho \beta ,\gamma } \\
&&)g_{\tau \lambda }g_{\mu \theta }g_{\nu \chi }g^{\alpha \beta }g^{\sigma
\gamma }g^{\rho \gamma ^{\prime }}(\delta _{\alpha }^{\tau }\delta _{\gamma
^{\prime }}^{\mu }\delta _{\sigma }^{\nu }+\delta _{\sigma }^{\tau }\delta
_{\alpha }^{\mu }\delta _{\gamma ^{\prime }}^{\nu }-\delta _{\gamma ^{\prime
}}^{\tau }\delta _{\alpha }^{\mu }\delta _{\sigma }^{\nu }) \\
&&-\frac{1}{4}g_{\tau \lambda }g_{\mu \theta }g_{\nu \chi }g^{\alpha \beta
}g^{\sigma \gamma }g^{\rho \gamma ^{\prime }}(\delta _{\rho }^{\tau }\delta
_{\gamma }^{\mu }\delta _{\sigma }^{\nu }+\delta _{\sigma }^{\tau }\delta
_{\rho }^{\mu }\delta _{\gamma }^{\nu }-\delta _{\gamma }^{\tau }\delta
_{\rho }^{\mu }\delta _{\sigma }^{\nu } \\
&&)(g_{\gamma ^{\prime }\beta ,\alpha }+g_{\alpha \gamma ^{\prime },\beta
}-g_{\alpha \beta ,\gamma ^{\prime }})-\frac{1}{4}(g_{\gamma \sigma ,\rho
}+g_{\rho \gamma ,\sigma }-g_{\rho \sigma ,\gamma }
\end{eqnarray*}

\begin{equation}
)g_{\tau \lambda }g_{\mu \theta }g_{\nu \chi }g^{\alpha \beta }g^{\sigma
\gamma }g^{\rho \gamma ^{\prime }}(\delta _{\alpha }^{\tau }\delta _{\gamma
^{\prime }}^{\mu }\delta _{\beta }^{\nu }+\delta _{\beta }^{\tau }\delta
_{\alpha }^{\mu }\delta _{\gamma ^{\prime }}^{\nu }-\delta _{\gamma ^{\prime
}}^{\tau }\delta _{\alpha }^{\mu }\delta _{\beta }^{\nu })].  \tag{2.12}
\end{equation}

\bigskip Using eq.(2.12), we finally achieve

\begin{equation*}
\pi _{\lambda \theta \chi }(\mathbf{x},t)=g_{\tau \lambda }g_{\mu \theta
}g_{\nu \chi }\pi ^{\tau \mu \nu }(\mathbf{x},t)=\kappa \lbrack \frac{-3}{2}%
g_{\chi \theta ,\lambda }+g_{\lambda \theta ,\chi }+\frac{1}{2}g_{\lambda
\chi }g_{\alpha \theta }^{,\alpha }+
\end{equation*}

\begin{equation}
\frac{3g_{\lambda \theta }g_{\alpha \chi }^{,\alpha }}{2}-g_{\theta \chi
}g_{\lambda \alpha }^{,\alpha }-\frac{g_{\lambda \chi }g^{\alpha \beta
}g_{\alpha \beta ,\theta }}{4}-\frac{3g_{\lambda \theta }g^{\alpha \beta
}g_{\alpha \beta ,\chi }}{4}+\frac{g_{\theta \chi }g^{\alpha \beta
}g_{\alpha \beta ,\lambda }}{2}].  \tag{2.13}
\end{equation}%
where the detail calculations see appendix A in supplied net material.

\section{III. Energy of gravitational field of the standard gravitational
Lagrangian}

\bigskip Using action (2.4) of gravitational field of the standard
gravtational Lagrangian
\begin{equation}
A=\int \mathcal{L}\sqrt{-g}dx^{4}=\int \kappa R\sqrt{-g}dx^{4},  \tag{3.1}
\end{equation}%
we get the energy of gravitational field of the standard gravtational
Lagrangian%
\begin{equation*}
H=\int (\pi ^{\mu \nu }(\mathbf{x},t)g_{\mu \nu ,t}(\mathbf{x},t)-\mathcal{L)%
}\sqrt{-g}dx^{4}=
\end{equation*}%
\begin{equation}
\int \kappa (\frac{g^{\alpha \beta }\partial R_{\alpha \beta }}{\partial
\partial _{t}g_{\mu \nu }(\mathbf{x},t)}g_{\mu \nu ,t}(\mathbf{x},t)-R%
\mathcal{)}\sqrt{-g}dx^{4}.  \tag{3.2}
\end{equation}

Eq.(3.2) means that if taking

\begin{equation}
\pi ^{\mu \nu }(\mathbf{x},t)]=\frac{\partial \mathcal{L}}{\partial \partial
_{t}g_{\mu \nu }(\mathbf{x},t)}=\kappa \sqrt{-g}\frac{\partial R}{\partial
\partial _{t}g_{\mu \nu }(\mathbf{x},t)},  \tag{3.3}
\end{equation}%
then the canonical momentum needs not to differentiate $\sqrt{-g}$ in
eq.(3.3), consequently, taking eq.(2.5) as the canonical momentum is
consistent and the most economic, or all momenta will be with $\sqrt{-g}$
througth the whole paper.

\bigskip Using eq.(2.13), we deduce three order contravariant tensor

\begin{equation*}
\pi ^{\gamma \alpha ^{\prime }\beta ^{\prime }}(\mathbf{x},t)=g^{\gamma
\lambda }g^{\alpha ^{\prime }\theta }g^{\beta ^{\prime }\chi }\pi _{\lambda
\theta \chi }(\mathbf{x},t)=g^{\gamma \lambda }g^{\alpha ^{\prime }\theta
}g^{\beta ^{\prime }\chi }g_{\tau \lambda }g_{\mu \theta }\cdot
\end{equation*}%
\begin{equation*}
g_{\nu \chi }\pi ^{\tau \mu \nu }(\mathbf{x},t)=\pi ^{\gamma \alpha ^{\prime
}\beta ^{\prime }}(\mathbf{x},t)=\kappa \lbrack \frac{-3}{2}g^{\gamma
\lambda }g^{\alpha ^{\prime }\theta }g^{\beta ^{\prime }\chi }g_{\chi \theta
,\lambda }+
\end{equation*}%
\begin{equation*}
g^{\gamma \lambda }g^{\alpha ^{\prime }\theta }g^{\beta ^{\prime }\chi
}g_{\lambda \theta ,\chi }+\frac{1}{2}g^{\gamma \lambda }g^{\alpha ^{\prime
}\theta }g^{\beta ^{\prime }\chi }g_{\lambda \chi }g_{\alpha \theta
}^{,\alpha }+\frac{3}{2}g^{\gamma \lambda }g^{\alpha ^{\prime }\theta
}g^{\beta ^{\prime }\chi }g_{\lambda \theta }g_{\alpha \chi }^{,\alpha }
\end{equation*}

\begin{equation*}
-g^{\gamma \lambda }g^{\alpha ^{\prime }\theta }g^{\beta ^{\prime }\chi
}g_{\theta \chi }g_{\lambda \alpha }^{,\alpha }-\frac{1}{4}g^{\gamma \lambda
}g^{\alpha ^{\prime }\theta }g^{\beta ^{\prime }\chi }g_{\lambda \chi
}g^{\alpha \beta }g_{\alpha \beta ,\theta }
\end{equation*}

\begin{equation*}
-\frac{3}{4}g^{\gamma \lambda }g^{\alpha ^{\prime }\theta }g^{\beta ^{\prime
}\chi }g_{\lambda \theta }g^{\alpha \beta }g_{\alpha \beta ,\chi }+\frac{1}{2%
}g^{\gamma \lambda }g^{\alpha ^{\prime }\theta }g^{\beta ^{\prime }\chi
}g_{\theta \chi }g^{\alpha \beta }g_{\alpha \beta ,\lambda }]=\kappa \lbrack
-
\end{equation*}

\begin{equation*}
\frac{3}{2}g^{\beta ^{\prime }\alpha ^{\prime },\gamma }+g^{\gamma \alpha
^{\prime },\beta ^{\prime }}+\frac{1}{2}g^{\alpha ^{\prime }\theta
}g^{\gamma \beta ^{\prime }}g_{\alpha \theta }^{,\alpha }+\frac{3}{2}%
g^{\beta ^{\prime }\chi }g_{\lambda \theta }^{\gamma \alpha ^{\prime
}}g_{\alpha \chi }^{,\alpha }-g^{\gamma \lambda }g^{\alpha ^{\prime }\beta
^{\prime }}g_{\lambda \alpha }^{,\alpha }
\end{equation*}

\begin{equation}
-\frac{1}{4}g^{\gamma \beta ^{\prime }}g^{\alpha \beta }g_{\alpha \beta
}^{,\alpha ^{\prime }}-\frac{3}{4}g^{\gamma \alpha ^{\prime }}g^{\alpha
\beta }g_{\alpha \beta }^{,\beta ^{\prime }}+\frac{1}{2}g^{\alpha ^{\prime
}\beta ^{\prime }}g^{\alpha \beta }g_{\alpha \beta }^{,\gamma }].  \tag{3.4}
\end{equation}

Because the index $\gamma $ is relevant to the time derivative, we take $%
\gamma $ as $t$ in eq.(3.4), then we get the two order contravariant tensor%
\begin{equation*}
\pi ^{\mu \nu }(\mathbf{x},t)\equiv \pi ^{t\mu \nu }(\mathbf{x},t)=\kappa (%
\frac{-3}{2}g^{\mu \nu ,t}+g^{t\mu ,\nu }
\end{equation*}

\begin{equation*}
+\frac{1}{2}g^{\mu \theta }g^{t\nu }g_{\alpha \theta }^{,\alpha }+\frac{3}{2}%
g^{\nu \chi }g^{t\mu }g_{\alpha \chi }^{,\alpha }-g^{t\lambda }g^{\mu \nu
}g_{\lambda \alpha }^{,\alpha }-
\end{equation*}

\begin{equation}
\frac{1}{4}g^{t\nu }g^{\alpha \beta }g_{\alpha \beta }^{,\mu }-\frac{3}{4}%
g^{t\mu }g^{\alpha \beta }g_{\alpha \beta }^{,\nu }+\frac{1}{2}g^{\mu \nu
}g^{\alpha \beta }g_{\alpha \beta }^{,t}).  \tag{3.5}
\end{equation}

Substituting eq.(3.5) into eq.(3.2), we deduce
\begin{equation*}
H=\int (\pi ^{\mu \nu }(\mathbf{x},t)g_{\mu \nu ,t}(\mathbf{x},t)-\mathcal{L)%
}\sqrt{-g}dx^{4}=
\end{equation*}%
\begin{equation*}
H=\int \kappa \lbrack (\frac{-3}{2}g^{\mu \nu ,t}+g^{t\mu ,\nu }+\frac{1}{2}%
g^{\mu \theta }g^{t\nu }g_{\alpha \theta }^{,\alpha }+
\end{equation*}

\begin{equation*}
\frac{3}{2}g^{\nu \chi }g^{t\mu }g_{\alpha \chi }^{,\alpha }-g^{t\lambda
}g^{\mu \nu }g_{\lambda \alpha }^{,\alpha }-\frac{1}{4}g^{t\nu }g^{\alpha
\beta }g_{\alpha \beta }^{,\mu }-\frac{3}{4}g^{t\mu }g^{\alpha \beta
}g_{\alpha \beta }^{,\nu }
\end{equation*}%
\begin{equation}
+\frac{1}{2}g^{\mu \nu }g^{\alpha \beta }g_{\alpha \beta }^{,t}))g_{\mu \nu
,t}(\mathbf{x},t)-g^{\alpha \beta }R_{\alpha \beta }]\sqrt{-g}dx^{4}.
\tag{3.6}
\end{equation}

It is very easy to calculate the standard gravitational system energy
eq.(3.6) when substituting concrete metric models into eq.(3.6).

\section{IV. Further investigations of quantization of quantum gravitational
fields}

\bigskip Because commutation relations of different fields and their
canonical momenta play very key role in quantization theory, we now study
the commutation relations.

For convenience and no losing generality, using eq.(3.5), we can define

\begin{equation*}
\pi ^{\mu \nu }(\mathbf{x},t)\equiv \pi ^{(0)\mu \nu }(\mathbf{x},t)+\pi
^{\prime (1)\mu \nu }(\mathbf{x},t)+\pi ^{\prime (2)\mu \nu }(\mathbf{x},t)
\end{equation*}%
\begin{equation*}
+\pi ^{\prime (3)\mu \nu }(\mathbf{x},t)=\pi ^{(3)\mu \nu }(\mathbf{x}%
,t)=\pi ^{\prime (3)\mu \nu }(\mathbf{x},t)+
\end{equation*}%
\begin{equation}
\pi ^{(2)\mu \nu }(\mathbf{x},t)=\pi ^{\prime (3)\mu \nu }(\mathbf{x},t)+\pi
^{\prime (2)\mu \nu }(\mathbf{x},t)+\pi ^{(1)\mu \nu }(\mathbf{x},t),
\tag{4.1}
\end{equation}%
where%
\begin{equation}
\pi ^{(0)\mu \nu }(\mathbf{x},t)=\kappa \frac{-3}{2}g^{\mu \nu ,t},
\tag{4.2}
\end{equation}

\begin{equation}
\pi ^{\prime (1)\mu \nu }(\mathbf{x},t)=\kappa g^{t\mu ,\nu },  \tag{4.3}
\end{equation}

\begin{equation}
\pi ^{\prime (2)\mu \nu }(\mathbf{x},t)=\kappa (\frac{1}{2}g^{\mu \theta
}g^{t\nu }g_{\alpha \theta }^{,\alpha }+\frac{3}{2}g^{\nu \chi }g^{t\mu
}g_{\alpha \chi }^{,\alpha }-g^{t\lambda }g^{\mu \nu }g_{\lambda \alpha
}^{,\alpha }),  \tag{4.4}
\end{equation}

\begin{equation}
\pi ^{\prime (3)\mu \nu }(\mathbf{x},t)=\kappa (\frac{1}{2}g^{\mu \nu
}g^{\alpha \beta }g_{\alpha \beta }^{,t}-\frac{1}{4}g^{t\nu }g^{\alpha \beta
}g_{\alpha \beta }^{,\mu }-\frac{3}{4}g^{t\mu }g^{\alpha \beta }g_{\alpha
\beta }^{,\nu }).  \tag{4.5}
\end{equation}

For $[\hat{g}_{\mu \nu }(\mathbf{x},t),\hat{\pi}^{(0)\alpha \beta }(\mathbf{x%
}^{\prime },t^{\prime })]_{t=t^{\prime }}$, we may define%
\begin{equation}
\hat{\pi}^{(0)\mu \nu }(\mathbf{x},t)=-i\frac{\partial }{\partial \hat{g}%
_{\mu \nu }(\mathbf{x},t)},  \tag{4.6}
\end{equation}%
then the classical Poisson bracket for any operators needs to be taken as%
\begin{equation*}
\{\hat{X}(\mathbf{x},t),\hat{Y}(\mathbf{x}^{\prime },t^{\prime
})\}_{pb,t=t^{\prime }}=
\end{equation*}%
\begin{equation}
\int (\frac{\partial \hat{X}(x)}{\partial \hat{g}_{\mu \nu }(y)}\frac{%
\partial \hat{Y}(x^{\prime })}{\partial \hat{\pi}^{(0)\mu \nu }(y)}-\frac{%
\partial \hat{X}(x)}{\partial \hat{\pi}^{(0)\mu \nu }(y)}\frac{\partial \hat{%
Y}(x^{\prime })}{\partial \hat{g}_{\mu \nu }(y)})_{pb,t=t^{\prime }}d\mathbf{%
y,}  \tag{4.7}
\end{equation}%
so that we can obtain the consistent theory between operator commutation
relations and classical Poisson brackets.

Therefore, we naturally have commutation relation of field operator and its
canonical conjugate momentum in the quantization for infinite degrees of
freedom%
\begin{equation*}
\frac{1}{i}[\hat{g}_{\mu \nu }(\mathbf{x},t),\hat{\pi}^{(0)\alpha \beta }(%
\mathbf{x}^{\prime },t^{\prime })]_{t=t^{\prime }}=\frac{1}{i}[\hat{g}_{\mu
\nu }(\mathbf{x},t),-i\frac{\partial }{\partial \hat{g}_{\alpha \beta }(%
\mathbf{x}^{\prime },t)}]_{t=t^{\prime }}
\end{equation*}%
\begin{equation*}
=\delta _{\mu }^{\alpha }\delta _{\nu }^{\beta }\delta (\mathbf{x-x}^{\prime
})=\Delta _{\mu \nu }^{(0)\alpha \beta }\delta (\mathbf{x-x}^{\prime })=-%
\hat{g}_{\mu \nu }(\mathbf{x},t)\frac{\partial }{\partial \hat{g}_{\alpha
\beta }(\mathbf{x},t)}
\end{equation*}%
\begin{equation*}
+\hat{g}_{\mu \nu }(\mathbf{x},t)\frac{\partial }{\partial \hat{g}_{\alpha
\beta }(\mathbf{x},t)}+\frac{\partial \hat{g}_{\mu \nu }(\mathbf{x},t)}{%
\partial \hat{g}_{\alpha \beta }(\mathbf{x},t)}=\int (
\end{equation*}%
\begin{equation*}
\frac{\partial \hat{g}_{\mu \nu }(\mathbf{x},t)}{\partial \hat{g}_{\rho
\sigma }(y)}\frac{\partial \hat{\pi}^{(0)\alpha \beta }(\mathbf{x}^{\prime
},t^{\prime })}{\partial \hat{\pi}^{(0)\rho \sigma }(y)}-\frac{\partial \hat{%
g}_{\mu \nu }(\mathbf{x},t)}{\partial \hat{\pi}^{(0)\rho \sigma }(y)}\frac{%
\partial \hat{\pi}^{(0)\alpha \beta }(\mathbf{x}^{\prime },t^{\prime })}{%
\partial \hat{g}_{\rho \sigma }(y)}
\end{equation*}%
\begin{equation}
)_{pb,t=t^{\prime }=t_{y}}d\mathbf{y}=\{\hat{g}_{\mu \nu }(\mathbf{x},t),%
\hat{\pi}^{(0)\alpha \beta }(\mathbf{x}^{\prime },t^{\prime
})\}_{pb,t=t^{\prime }},  \tag{4.8}
\end{equation}%
where $\{\hat{g}_{\mu \nu }(\mathbf{x},t),\hat{\pi}^{(0)\alpha \beta }(%
\mathbf{x}^{\prime },t^{\prime })\}_{pb,t=t^{\prime }}$ is classical Poisson
bracket for infinite degrees of freedom.

For infinite degrees of freedom and similar to investigations on Eq.(4.8),
we have%
\begin{equation*}
\frac{1}{i}[\hat{g}_{\mu \nu }(\mathbf{x},t),\hat{g}_{\alpha \beta }(\mathbf{%
x}^{\prime },t)]_{t=t^{\prime }}=\frac{1}{i}[\hat{g}_{\mu \nu }(\mathbf{x},t)%
\hat{g}_{\alpha \beta }(\mathbf{x}^{\prime },t)-
\end{equation*}%
\begin{equation*}
\hat{g}_{\alpha \beta }(\mathbf{x}^{\prime },t)\hat{g}_{\mu \nu }(\mathbf{x}%
,t)]_{t=t^{\prime }}=0=\int (\frac{\partial \hat{g}_{\mu \nu }(\mathbf{x},t)%
}{\partial \hat{g}_{\rho \sigma }(y)}\frac{\partial \hat{g}_{\alpha \beta }(%
\mathbf{x}^{\prime },t^{\prime })}{\partial \hat{\pi}^{(0)\rho \sigma }(y)}-
\end{equation*}%
\begin{equation}
\frac{\partial \hat{g}_{\mu \nu }(\mathbf{x},t)}{\partial \hat{\pi}^{(0)\rho
\sigma }(y)}\frac{\partial \hat{g}_{\alpha \beta }(\mathbf{x}^{\prime
},t^{\prime })}{\partial \hat{g}_{\rho \sigma }(y)})_{pb,t=t^{\prime
}=t_{y}}d\mathbf{y}=\{\hat{g}_{\mu \nu }(\mathbf{x},t),\hat{g}_{\alpha \beta
}(\mathbf{x}^{\prime },t^{\prime })\}_{pb,t=t^{\prime }}  \tag{4.9}
\end{equation}

\begin{equation*}
\frac{1}{i}[\hat{\pi}^{(0)\mu \nu }(\mathbf{x},t),\hat{\pi}^{(0)\alpha \beta
}(\mathbf{x}^{\prime },t^{\prime })]_{t=t^{\prime }}=\frac{1}{i}[\hat{\pi}%
^{(0)\mu \nu }(\mathbf{x},t)\hat{\pi}^{(0)\alpha \beta }(\mathbf{x}^{\prime
},t^{\prime })
\end{equation*}%
\begin{equation*}
-\hat{\pi}^{(0)\alpha \beta }(\mathbf{x}^{\prime },t^{\prime })\hat{\pi}%
^{(0)\mu \nu }(\mathbf{x},t)]_{t=t^{\prime }}=0=\int (
\end{equation*}%
\begin{equation*}
\frac{\partial \hat{\pi}^{(0)\mu \nu }(\mathbf{x},t)}{\partial \hat{g}_{\rho
\sigma }(y)}\frac{\partial \hat{\pi}^{(0)\alpha \beta }(\mathbf{x}^{\prime
},t^{\prime })}{\partial \hat{\pi}^{(0)\rho \sigma }(y)}-\frac{\partial \hat{%
\pi}^{(0)\mu \nu }(\mathbf{x},t)}{\partial \hat{\pi}^{(0)\rho \sigma }(y)}%
\frac{\partial \hat{\pi}^{(0)\alpha \beta }(\mathbf{x}^{\prime },t^{\prime })%
}{\partial \hat{g}_{\rho \sigma }(y)}
\end{equation*}%
\begin{equation}
)_{pb,t=t^{\prime }=t_{y}}d\mathbf{y}=\{\hat{\pi}^{(0)\mu \nu }(\mathbf{x}%
,t),\hat{\pi}^{(0)\alpha \beta }(\mathbf{x}^{\prime },t^{\prime
})\}_{pb,t=t^{\prime }}.  \tag{4.10}
\end{equation}

Using the investigations of this section, we can give many important
investigations. In terms of the detailed argument in this section, we do
find that, for the commutative relations of operators, people can do exact
calculations directly with the classical Poisson's bracket of operators
because they are completely equivalent by the relations above.

\section{V. Commutation relation for gravitational fields and the first
style of momenta}

We further consider the commutation relation for gravitational fields and
the first style of momenta $\hat{\pi}^{(1)\alpha \beta }(\mathbf{x}^{\prime
},t^{\prime })$

\begin{equation*}
\frac{1}{i}[\hat{g}_{\mu \nu }(\mathbf{x},t),\hat{\pi}^{(1)\alpha \beta }(%
\mathbf{x}^{\prime },t^{\prime })]_{t=t^{\prime }}=\frac{1}{i}[\hat{g}_{\mu
\nu }(\mathbf{x},t),\hat{\pi}^{(0)\alpha \beta }(\mathbf{x}^{\prime
},t^{\prime })]_{t=t^{\prime }}
\end{equation*}%
\begin{equation*}
+\frac{1}{i}[\hat{g}_{\mu \nu }(\mathbf{x},t),\hat{\pi}^{\prime (1)\alpha
\beta }(\mathbf{x}^{\prime },t^{\prime })]_{t=t^{\prime }}=\frac{1}{i}[\hat{g%
}_{\mu \nu }(\mathbf{x},t),\hat{\pi}^{(0)\alpha \beta }(\mathbf{x}^{\prime
},t^{\prime })
\end{equation*}%
\begin{equation*}
]_{t=t^{\prime }}+\frac{1}{i}[\hat{g}_{\mu \nu }(\mathbf{x},t),\kappa
g^{t\alpha ,\beta }]_{t=t^{\prime }}=\frac{1}{i}[\hat{g}_{\mu \nu }(\mathbf{x%
},t),\hat{\pi}^{(0)\alpha \beta }(\mathbf{x}^{\prime },t^{\prime
})]_{t=t^{\prime }}+
\end{equation*}%
\begin{equation*}
\int (\frac{\partial \hat{g}_{\mu \nu }(\mathbf{x},t)}{\partial \hat{g}%
_{\rho \sigma }(y)}\frac{\partial (\kappa g^{t\alpha ,\beta })}{\partial
(\kappa \frac{-3}{2}g^{\rho \sigma ,t})}-\frac{\partial \hat{g}_{\mu \nu }(%
\mathbf{x},t)}{\partial \hat{\pi}^{(0)\rho \sigma }(y)}\frac{\partial \hat{%
\pi}^{\prime (1)\alpha \beta }(\mathbf{x}^{\prime },t^{\prime })}{\partial
\hat{g}_{\rho \sigma }(y)}
\end{equation*}%
\begin{equation}
)_{pb,t=t^{\prime }=t_{y}}d\mathbf{y}=\{\hat{g}_{\mu \nu }(\mathbf{x},t),%
\hat{\pi}^{(1)\alpha \beta }(\mathbf{x}^{\prime },t^{\prime
})\}_{pb,t=t^{\prime }}.  \tag{5.1}
\end{equation}%
Thus we can further deduce%
\begin{equation*}
\frac{1}{i}[\hat{g}_{\mu \nu }(\mathbf{x},t),\hat{\pi}^{(1)\alpha \beta }(%
\mathbf{x}^{\prime },t^{\prime })]_{t=t^{\prime }}=\delta _{\mu }^{\alpha
}\delta _{\nu }^{\beta }\delta (\mathbf{x-x}^{\prime })
\end{equation*}%
\begin{equation*}
-\frac{2}{3}\int (\delta _{\mu }^{\rho }\delta _{\nu }^{\sigma }\delta (%
\mathbf{x-y})\delta _{\rho }^{t}\delta _{\sigma }^{\alpha }\delta
_{t}^{\beta }\delta (\mathbf{x}^{\prime }\mathbf{-y}))_{pb,t=t^{\prime
}=t_{y}}d\mathbf{y}
\end{equation*}

\begin{equation*}
=\delta _{\mu }^{\alpha }\delta _{\nu }^{\beta }\delta (\mathbf{x-x}^{\prime
})-\frac{2}{3}\delta _{\mu }^{t}\delta _{\nu }^{\alpha }\delta (\mathbf{x-x}%
^{\prime })\delta _{t}^{\beta }
\end{equation*}%
\begin{equation}
=(\delta _{\mu }^{\alpha }\delta _{\nu }^{\beta }-\frac{2}{3}\delta _{\mu
}^{\beta }\delta _{\nu }^{\alpha })\delta (\mathbf{x-x}^{\prime })=\Delta
_{\mu \nu }^{(1)\alpha \beta }\delta (\mathbf{x-x}^{\prime }),  \tag{5.2}
\end{equation}%
where the first term and the second term on right side of eq.(5.2) are
linear terms, which shows the only linear property of the standard
gravitational Lagrangian, the nonlinear property will be shown in the
following higher order term studies.

\section{VI. Commutation relation for gravitational fields and the second
style of momenta}

\bigskip We now consider the commutation relation for gravitational fields
and the second style of momenta $\hat{\pi}^{(2)\alpha \beta }(\mathbf{x}%
^{\prime },t^{\prime })$

\begin{equation*}
\frac{1}{i}[\hat{g}_{\mu ^{\prime }\nu ^{\prime }}(\mathbf{x},t),\hat{\pi}%
^{(2)\mu \nu }(\mathbf{x}^{\prime },t^{\prime })]_{t=t^{\prime }}=\frac{1}{i}%
[\hat{g}_{\mu ^{\prime }\nu ^{\prime }}(\mathbf{x},t),\hat{\pi}^{(1)\mu \nu
}(\mathbf{x}^{\prime },t^{\prime })
\end{equation*}%
\begin{equation*}
]_{t=t^{\prime }}+\frac{1}{i}[\hat{g}_{\mu ^{\prime }\nu ^{\prime }}(\mathbf{%
x},t),\hat{\pi}^{\prime (2)\mu \nu }(\mathbf{x}^{\prime },t^{\prime
})]_{t=t^{\prime }}=\frac{1}{i}[\hat{g}_{\mu ^{\prime }\nu ^{\prime }}(%
\mathbf{x},t),\hat{\pi}^{(1)\mu \nu }(
\end{equation*}%
\begin{equation*}
\mathbf{x}^{\prime },t^{\prime })]_{t=t^{\prime }}+\frac{1}{i}[\hat{g}_{\mu
^{\prime }\nu ^{\prime }}(\mathbf{x},t),\kappa (\frac{1}{2}g^{\mu \theta
}g^{t\nu }g_{\alpha \theta }^{,\alpha }+\frac{3}{2}g^{\nu \chi }g^{t\mu
}g_{\alpha \chi }^{,\alpha }-
\end{equation*}%
\begin{equation*}
g^{t\lambda }g^{\mu \nu }g_{\lambda \alpha }^{,\alpha })]_{t=t^{\prime }}=%
\frac{[\hat{g}_{\mu ^{\prime }\nu ^{\prime }}(\mathbf{x},t),\hat{\pi}%
^{(1)\mu \nu }(\mathbf{x}^{\prime },t^{\prime })]_{t=t^{\prime }}}{i}+\int (%
\frac{\partial \hat{g}_{\mu ^{\prime }\nu ^{\prime }}(\mathbf{x},t)}{%
\partial \hat{g}_{\rho \sigma }(y)}
\end{equation*}%
\begin{equation*}
\cdot \frac{\partial \kappa (\frac{1}{2}g^{\mu \theta }g^{t\nu }g^{\alpha
\gamma }g_{\alpha \theta ,\gamma }+\frac{3}{2}g^{\nu \chi }g^{t\mu
}g^{\alpha \gamma }g_{\alpha \chi ,\gamma }-g^{t\lambda }g^{\mu \nu
}g^{\alpha \gamma }g_{\lambda \alpha ,\gamma })}{\partial (\kappa \frac{-3}{2%
}g^{\rho \sigma ,t})}
\end{equation*}%
\begin{equation*}
)_{pb,t=t^{\prime }=t_{y}}d\mathbf{y}=\{\hat{g}_{\mu ^{\prime }\nu ^{\prime
}}(\mathbf{x},t),\hat{\pi}^{(2)\mu \nu }(\mathbf{x}^{\prime },t^{\prime
})\}_{pb,t=t^{\prime }}=\frac{1}{i}[
\end{equation*}%
\begin{equation*}
\hat{g}_{\mu ^{\prime }\nu ^{\prime }}(\mathbf{x},t),\hat{\pi}^{(1)\mu \nu }(%
\mathbf{x}^{\prime },t^{\prime })]_{t=t^{\prime }}+\int d\mathbf{y}(\frac{%
\partial \hat{g}_{\mu ^{\prime }\nu ^{\prime }}(\mathbf{x},t)}{\partial \hat{%
g}_{\rho \sigma }(y)}\frac{\partial }{\partial (\kappa \frac{-3}{2}g^{\rho
\sigma ,t})}
\end{equation*}%
\begin{equation}
\kappa (\frac{g^{\mu \theta }g^{t\nu }g^{\alpha \gamma }g_{\alpha \theta
,\gamma }}{2}+\frac{3}{2}g^{\nu \chi }g^{t\mu }g^{\alpha \gamma }g_{\alpha
\chi ,\gamma }-g^{t\lambda }g^{\mu \nu }g^{\alpha \gamma }g_{\lambda \alpha
,\gamma }))_{t=t^{\prime }=t_{y}}.  \tag{6.1}
\end{equation}%
Using%
\begin{equation}
g^{\sigma \mu }g_{\mu \beta ,\rho }=-g_{,\rho }^{\sigma \mu }g_{\mu \beta
},g_{\alpha \sigma }g^{\sigma \mu }g_{\mu \beta ,\rho }=-g_{\alpha \sigma
}g_{,\rho }^{\sigma \mu }g_{\mu \beta }=g_{\alpha \beta ,\rho },  \tag{6.2}
\end{equation}%
we can have%
\begin{equation*}
\hat{\pi}^{\prime (2)\mu \nu }(\mathbf{x}^{\prime },t^{\prime })=\kappa (%
\frac{1}{2}g^{\mu \theta }g^{t\nu }g^{\alpha \gamma }g_{\alpha \theta
,\gamma }+\frac{3}{2}g^{\nu \chi }g^{t\mu }g^{\alpha \gamma }g_{\alpha \chi
,\gamma }-
\end{equation*}%
\begin{equation*}
g^{t\lambda }g^{\mu \nu }g^{\alpha \gamma }g_{\lambda \alpha ,\gamma
})=\kappa (-\frac{g^{\mu \theta }g^{t\nu }g^{\alpha \gamma }g_{\alpha \sigma
}g_{,\gamma }^{\sigma \mu ^{\prime }}g_{\mu ^{\prime }\theta }}{2}-\frac{%
3g^{\nu \chi }g^{t\mu }g^{\alpha \gamma }(}{2}
\end{equation*}%
\begin{equation*}
g_{\alpha \sigma }g_{,\gamma }^{\sigma \mu ^{\prime }}g_{\mu ^{\prime }\chi
})-g^{t\lambda }g^{\mu \nu }g^{\alpha \gamma }(-g_{\lambda \mu ^{\prime
}}g_{,\gamma }^{\mu ^{\prime }\sigma }g_{\sigma \alpha }))=\kappa (\frac{1}{2%
}g^{\mu \theta }g^{t\nu }(-
\end{equation*}%
\begin{equation*}
g_{\alpha \sigma }g^{\sigma \mu ^{\prime },\alpha }g_{\mu ^{\prime }\theta
})+\frac{3}{2}g^{\nu \chi }g^{t\mu }(-g_{\alpha \sigma }g^{\sigma \mu
^{\prime },\alpha }g_{\mu ^{\prime }\chi })-g^{t\lambda }g^{\mu \nu }(-
\end{equation*}%
\begin{equation}
g_{\lambda \mu ^{\prime }}g^{\mu ^{\prime }\sigma ,\alpha }g_{\sigma \alpha
}))=\kappa (\frac{1}{2}g^{t\nu }(-g_{\alpha \sigma }g^{\sigma \mu ,\alpha })+%
\frac{3}{2}g^{t\mu }(-g_{\alpha \sigma }g^{\sigma \nu ,\alpha })-g^{\mu \nu
}(-g^{t\sigma ,\alpha }g_{\sigma \alpha })).  \tag{6.3}
\end{equation}%
Putting eq.(6.3) into eq.(6.1), we have

\begin{equation*}
\frac{1}{i}[\hat{g}_{\mu ^{\prime }\nu ^{\prime }}(\mathbf{x},t),\hat{\pi}%
^{(2)\mu \nu }(\mathbf{x}^{\prime },t^{\prime })]_{t=t^{\prime }}=\frac{1}{i}%
[\hat{g}_{\mu ^{\prime }\nu ^{\prime }}(\mathbf{x},t),\hat{\pi}^{(1)\mu \nu
}(\mathbf{x}^{\prime },t^{\prime })]_{t=t^{\prime }}
\end{equation*}%
\begin{equation*}
+\kappa \int d\mathbf{y}(\frac{\partial \hat{g}_{\mu ^{\prime }\nu ^{\prime
}}(\mathbf{x},t)}{\partial \hat{g}_{\rho \sigma ^{\prime }}(y)}\frac{%
\partial }{\partial (\kappa \frac{-3}{2}g^{\rho \sigma ^{\prime },t})}(\frac{%
1}{2}g^{t\nu }(-g_{\alpha \sigma }g^{\sigma \mu ,\alpha })
\end{equation*}%
\begin{equation*}
+\frac{3}{2}g^{t\mu }(-g_{\alpha \sigma }g^{\sigma \nu ,\alpha })-g^{\mu \nu
}(-g^{t\sigma ,\alpha }g_{\sigma \alpha }))_{pb,t=t^{\prime }=t_{y}}
\end{equation*}%
\begin{equation*}
=\frac{1}{i}[\hat{g}_{\mu ^{\prime }\nu ^{\prime }}(\mathbf{x},t),\hat{\pi}%
^{(1)\mu \nu }(\mathbf{x}^{\prime },t^{\prime })]_{t=t^{\prime }}+\int d%
\mathbf{y}(\delta _{\mu ^{\prime }}^{\rho }\delta _{\nu ^{\prime }}^{\sigma
^{\prime }}\delta (\mathbf{x-y)}(-
\end{equation*}%
\begin{equation*}
\frac{g^{t\nu }g_{\alpha \sigma }\delta _{\rho }^{\sigma }\delta _{\sigma
^{\prime }}^{\mu }\delta _{t}^{\alpha }}{2}+\frac{3g^{t\mu }(-g_{\alpha
\sigma }\delta _{\rho }^{\sigma }\delta _{\sigma ^{\prime }}^{\nu }\delta
_{t}^{\alpha })}{2}+
\end{equation*}%
\begin{equation*}
g^{\mu \nu }\delta _{\rho }^{t}\delta _{\sigma ^{\prime }}^{\sigma }\delta
_{t}^{\alpha }g_{\sigma \alpha })\delta (\mathbf{y-x}^{\prime }\mathbf{)}=%
\frac{1}{i}[\hat{g}_{\mu ^{\prime }\nu ^{\prime }}(\mathbf{x},t),\hat{\pi}%
^{(1)\mu \nu }(\mathbf{x}^{\prime },t^{\prime })]_{t=t^{\prime }}+
\end{equation*}%
\begin{equation*}
(\delta _{\mu ^{\prime }}^{\rho }\delta _{\nu ^{\prime }}^{\sigma ^{\prime
}}\kappa (\frac{1}{2}g^{t\nu }(-g_{\alpha \sigma }\delta _{\rho }^{\sigma
}\delta _{\sigma ^{\prime }}^{\mu }\delta _{t}^{\alpha })+\frac{3}{2}g^{t\mu
}(-g_{\alpha \sigma }\delta _{\rho }^{\sigma }\delta _{\sigma ^{\prime
}}^{\nu }\delta _{t}^{\alpha })-
\end{equation*}%
\begin{equation*}
g^{\mu \nu }(-\delta _{\rho }^{t}\delta _{\sigma ^{\prime }}^{\sigma }\delta
_{t}^{\alpha }g_{\sigma \alpha }))\delta (\mathbf{x-x}^{\prime }\mathbf{)}=%
\frac{1}{i}[
\end{equation*}%
\begin{equation*}
\hat{g}_{\mu ^{\prime }\nu ^{\prime }}(\mathbf{x},t),\hat{\pi}^{(1)\mu \nu }(%
\mathbf{x}^{\prime },t^{\prime })]_{t=t^{\prime }}+((\frac{1}{2}g^{t\nu
}(-g_{t\mu ^{\prime }}\delta _{\nu ^{\prime }}^{\mu })+\frac{3}{2}g^{t\mu
}(-g_{t\mu ^{\prime }}\delta _{\nu ^{\prime }}^{\nu })
\end{equation*}

\begin{equation*}
-g^{\mu \nu }(-\delta _{\mu ^{\prime }}^{t}g_{\nu ^{\prime }t}))\delta (%
\mathbf{x-x}^{\prime }\mathbf{)}=\frac{1}{i}[\hat{g}_{\mu ^{\prime }\nu
^{\prime }}(\mathbf{x},t),\hat{\pi}^{(1)\mu \nu }(\mathbf{x}^{\prime
},t^{\prime })]_{t=t^{\prime }}
\end{equation*}%
\begin{equation}
+((-\frac{1}{2}\delta _{\mu ^{\prime }}^{\nu }\delta _{\nu ^{\prime }}^{\mu
})-\frac{3}{2}\delta _{\mu ^{\prime }}^{\mu }\delta _{\nu ^{\prime }}^{\nu
}-g^{\mu \nu }g_{\nu ^{\prime }\mu ^{\prime }})\delta (\mathbf{x-x}^{\prime }%
\mathbf{).}  \tag{6.4}
\end{equation}

Putting eq.(5.2) into eq.(6.4), it follows that

\begin{equation*}
\frac{1}{i}[\hat{g}_{\mu ^{\prime }\nu ^{\prime }}(\mathbf{x},t),\hat{\pi}%
^{(2)\mu \nu }(\mathbf{x}^{\prime },t^{\prime })]_{t=t^{\prime }}=(\delta
_{\mu ^{\prime }}^{\mu }\delta _{\nu ^{\prime }}^{\nu }-\frac{2}{3}\delta
_{\mu ^{\prime }}^{\nu }\delta _{\nu ^{\prime }}^{\mu })\delta (\mathbf{x-x}%
^{\prime })
\end{equation*}%
\begin{equation*}
+((-\frac{1}{2}\delta _{\mu ^{\prime }}^{\nu }\delta _{\nu ^{\prime }}^{\mu
})-\frac{3}{2}\delta _{\mu ^{\prime }}^{\mu }\delta _{\nu ^{\prime }}^{\nu
}-g^{\mu \nu }g_{\nu ^{\prime }\mu ^{\prime }})\delta (\mathbf{x-x}^{\prime }%
\mathbf{)}=-(
\end{equation*}

\begin{equation}
\frac{1}{2}\delta _{\mu ^{\prime }}^{\nu }\delta _{\nu ^{\prime }}^{\mu }+%
\frac{7}{6}\delta _{\mu ^{\prime }}^{\mu }\delta _{\nu ^{\prime }}^{\nu
}+g^{\mu \nu }g_{\nu ^{\prime }\mu ^{\prime }})\delta (\mathbf{x-x}^{\prime }%
\mathbf{)}=\Delta _{\mu ^{\prime }\nu ^{\prime }}^{(2)\mu \nu }\delta (%
\mathbf{x-x}^{\prime }\mathbf{),}  \tag{6.5}
\end{equation}%
where the first term and the second term on right side of eq.(6.5) are
linear terms, the third term is nonlinear term, which shows the nonlinear
property of the nonlinear system of the standard gravitational Lagrangian.

\section{VII. Commutation relation for graviational fields and the third
style of momenta}

We now consider the commutation relation for gravitational fields and the
third style of momenta $\hat{\pi}^{(3)\alpha \beta }(\mathbf{x}^{\prime
},t^{\prime })$

\begin{equation*}
\frac{1}{i}[\hat{g}_{\mu ^{\prime }\nu ^{\prime }}(\mathbf{x},t),\hat{\pi}%
^{(3)\mu \nu }(\mathbf{x}^{\prime },t^{\prime })]_{t=t^{\prime }}=\frac{1}{i}%
[\hat{g}_{\mu ^{\prime }\nu ^{\prime }}(\mathbf{x},t),\hat{\pi}^{(2)\mu \nu
}(\mathbf{x}^{\prime },t^{\prime })
\end{equation*}%
\begin{equation*}
]_{t=t^{\prime }}+\frac{1}{i}[\hat{g}_{\mu ^{\prime }\nu ^{\prime }}(\mathbf{%
x},t),\hat{\pi}^{\prime (3)\mu \nu }(\mathbf{x}^{\prime },t^{\prime
})]_{t=t^{\prime }}=\frac{1}{i}[\hat{g}_{\mu ^{\prime }\nu ^{\prime }}(%
\mathbf{x},t),\hat{\pi}^{(2)\mu \nu }(
\end{equation*}%
\begin{equation*}
\mathbf{x}^{\prime },t^{\prime })]_{t=t^{\prime }}+\frac{1}{i}[\hat{g}_{\mu
^{\prime }\nu ^{\prime }}(\mathbf{x},t),\kappa (\frac{1}{2}g^{\mu \nu
}g^{\alpha \beta }g_{\alpha \beta }^{,t}-\frac{1}{4}g^{t\nu }g^{\alpha \beta
}g_{\alpha \beta }^{,\mu }
\end{equation*}%
\begin{equation*}
-\frac{3}{4}g^{t\mu }g^{\alpha \beta }g_{\alpha \beta }^{,\nu
})]_{t=t^{\prime }}=\frac{1}{i}[\hat{g}_{\mu ^{\prime }\nu ^{\prime }}(%
\mathbf{x},t),\hat{\pi}^{(2)\mu \nu }(\mathbf{x}^{\prime },t^{\prime
})]_{t=t^{\prime }}+
\end{equation*}%
\begin{equation*}
\kappa \int (\frac{\partial \hat{g}_{\mu ^{\prime }\nu ^{\prime }}(\mathbf{x}%
,t)}{\partial \hat{g}_{\rho \sigma }(y)}\frac{\partial (\frac{1}{2}g^{\mu
\nu }g^{\alpha \beta }g_{\alpha \beta }^{,t}-\frac{1}{4}g^{t\nu }g^{\alpha
\beta }g_{\alpha \beta }^{,\mu }-\frac{3}{4}g^{t\mu }g^{\alpha \beta
}g_{\alpha \beta }^{,\nu })}{\partial (\kappa \frac{-3}{2}g^{\rho \sigma ,t})%
}
\end{equation*}%
\begin{equation}
)_{pb,t=t^{\prime }=t_{y}}d\mathbf{y}=\{\hat{g}_{\mu ^{\prime }\nu ^{\prime
}}(\mathbf{x},t),\hat{\pi}^{(3)\mu \nu }(\mathbf{x}^{\prime },t^{\prime
})\}_{pb,t=t^{\prime }}.  \tag{7.1}
\end{equation}%
Using eq.(4.5), we have

\begin{equation*}
\pi ^{\prime (3)\mu \nu }(\mathbf{x},t)=\kappa (\frac{1}{2}g^{\mu \nu
}g^{\alpha \beta }g_{\alpha \beta }^{,t}-\frac{1}{4}g^{t\nu }g^{\alpha \beta
}g_{\alpha \beta }^{,\mu }-\frac{3}{4}g^{t\mu }g^{\alpha \beta }g_{\alpha
\beta }^{,\nu }).
\end{equation*}%
\begin{eqnarray*}
&=&\kappa (\frac{1}{2}g^{\mu \nu }g^{\alpha \beta }g^{t\gamma }(-g_{\alpha
\sigma }g_{,\gamma }^{\sigma \varepsilon }g_{\varepsilon \beta })-\frac{1}{4}%
g^{t\nu }g^{\alpha \beta }g^{\mu \gamma }(-g_{\alpha \sigma }g_{,\gamma
}^{\sigma \varepsilon }g_{\varepsilon \beta }) \\
&&-\frac{3}{4}g^{t\mu }g^{\alpha \beta }g^{\nu \gamma }(-g_{\alpha \sigma
}g_{,\gamma }^{\sigma \varepsilon }g_{\varepsilon \beta }))
\end{eqnarray*}

\begin{equation}
=\kappa (\frac{1}{2}g^{\mu \nu }(-g^{\beta \varepsilon ,t}g_{\varepsilon
\beta })-\frac{1}{4}g^{t\nu }(-g^{\beta \varepsilon ,\mu }g_{\varepsilon
\beta })-\frac{3}{4}g^{t\mu }(-g^{\beta \varepsilon ,\nu }g_{\varepsilon
\beta })).  \tag{7.2}
\end{equation}

Putting eq.(7.2) into eq.(7.1), we deduce

\begin{equation*}
\frac{1}{i}[\hat{g}_{\mu ^{\prime }\nu ^{\prime }}(\mathbf{x},t),\hat{\pi}%
^{(3)\mu \nu }(\mathbf{x}^{\prime },t^{\prime })]_{t=t^{\prime }}=\frac{1}{i}%
[\hat{g}_{\mu ^{\prime }\nu ^{\prime }}(\mathbf{x},t),\hat{\pi}^{(2)\mu \nu
}(\mathbf{x}^{\prime },t^{\prime })
\end{equation*}%
\begin{equation*}
]_{t=t^{\prime }}+\int d\mathbf{y}(\frac{\partial \hat{g}_{\mu ^{\prime }\nu
^{\prime }}(\mathbf{x},t)}{\partial \hat{g}_{\rho \sigma }(y)}\frac{\partial
}{\partial (\kappa \frac{-3}{2}g^{\rho \sigma ,t})}\kappa (\frac{1}{2}g^{\mu
\nu }(-g^{\beta \varepsilon ,t}g_{\varepsilon \beta })
\end{equation*}%
\begin{equation*}
-\frac{1}{4}g^{t\nu }(-g^{\beta \varepsilon ,\mu }g_{\varepsilon \beta })-%
\frac{3}{4}g^{t\mu }(-g^{\beta \varepsilon ,\nu }g_{\varepsilon \beta
})))_{pb,t=t^{\prime }=t_{y}}
\end{equation*}

\begin{equation*}
=\frac{1}{i}[\hat{g}_{\mu ^{\prime }\nu ^{\prime }}(\mathbf{x},t),\hat{\pi}%
^{(2)\mu \nu }(\mathbf{x}^{\prime },t^{\prime })]_{t=t^{\prime }}+\int d%
\mathbf{y}(\delta _{\mu ^{\prime }}^{\rho }\delta _{\nu ^{\prime }}^{\sigma
}\delta (\mathbf{x-y)}\cdot
\end{equation*}%
\begin{equation*}
(\frac{2}{3}g^{\mu \nu }(\delta _{\rho }^{\beta }\delta _{\sigma
}^{\varepsilon }\delta _{t}^{t}g_{\varepsilon \beta })+\frac{1}{6}g^{t\nu
}(-\delta _{\rho }^{\beta }\delta _{\sigma }^{\varepsilon }\delta _{t}^{\mu
}g_{\varepsilon \beta })+\frac{1}{2}g^{t\mu }(-\delta _{\rho }^{\beta
}\delta _{\sigma }^{\varepsilon }\delta _{t}^{\nu }g_{\varepsilon \beta })
\end{equation*}

\begin{equation*}
)_{pb,t=t^{\prime }=t_{y}}=\frac{1}{i}[\hat{g}_{\mu ^{\prime }\nu ^{\prime
}}(\mathbf{x},t),\hat{\pi}^{(2)\mu \nu }(\mathbf{x}^{\prime },t^{\prime
})]_{t=t^{\prime }}+\int d\mathbf{y}(\delta (\mathbf{x-y)}\cdot
\end{equation*}%
\begin{equation*}
(\frac{2}{3}g^{\mu \nu }g_{\nu ^{\prime }\mu ^{\prime }}+\frac{1}{6}g^{\mu
\nu }(-g_{\nu ^{\prime }\mu ^{\prime }})+\frac{1}{2}g^{\nu \mu }(-g_{\nu
^{\prime }\mu ^{\prime }}))_{pb,t=t^{\prime }=t_{y}}\delta (\mathbf{x}%
^{\prime }\mathbf{-y)}
\end{equation*}

\begin{equation*}
=\frac{1}{i}[\hat{g}_{\mu ^{\prime }\nu ^{\prime }}(\mathbf{x},t),\hat{\pi}%
^{(2)\mu \nu }(\mathbf{x}^{\prime },t^{\prime })]_{t=t^{\prime }}+(\delta (%
\mathbf{x-x}^{\prime }\mathbf{)}\cdot (0))_{pb,t=t^{\prime }=t_{y}}=
\end{equation*}%
\begin{equation}
-(\frac{1}{2}\delta _{\mu ^{\prime }}^{\nu }\delta _{\nu ^{\prime }}^{\mu }+%
\frac{7}{6}\delta _{\mu ^{\prime }}^{\mu }\delta _{\nu ^{\prime }}^{\nu
}+g^{\mu \nu }g_{\nu ^{\prime }\mu ^{\prime }})\delta (\mathbf{x-x}^{\prime }%
\mathbf{)=}\Delta _{\mu ^{\prime }\nu ^{\prime }}^{(3)\mu \nu }\delta (%
\mathbf{x-x}^{\prime }\mathbf{).}  \tag{7.3}
\end{equation}

It is interesting that eq.(7.3) has nothing to do with $\hat{\pi}^{\prime
(3)\mu \nu }(\mathbf{x}^{\prime },t^{\prime })$, and is only relevant to $%
\hat{\pi}^{(2)\mu \nu }(\mathbf{x}^{\prime },t^{\prime })$, these charactors
just reflect the symmetric property $\Delta _{\mu \nu }^{(3)\alpha \beta
}=\Delta _{\mu \nu }^{(2)\alpha \beta }$ of the nonlinear system of the
standard gravitational Lagrangian.\bigskip

Using eq.(7.3) and taking $(\mu ^{\prime },\nu ^{\prime })=(\mu ,\nu )$, we
have

\begin{equation*}
\frac{1}{i}[\hat{g}_{\mu \nu }(\mathbf{x},t),\hat{\pi}^{(3)\mu \nu }(\mathbf{%
x}^{\prime },t^{\prime })]_{t=t^{\prime }}=-(\frac{1}{2}\delta _{\mu }^{\nu
}\delta _{\nu }^{\mu }+\frac{7}{6}\delta _{\mu }^{\mu }\delta _{\nu }^{\nu
}+g^{\mu \nu }g_{\nu \mu }
\end{equation*}

\begin{equation}
)\delta (\mathbf{x-x}^{\prime }\mathbf{)=-(\frac{2}{3}+24)\delta (\mathbf{x-x%
}^{\prime }\mathbf{)}=}\Delta _{\mu \nu }^{(3)\mu \nu }\delta (\mathbf{x-x}%
^{\prime }\mathbf{)}  \tag{7.4}
\end{equation}%
When the repeating indexes don't sum up, eq.(7.4) is

\begin{equation*}
\frac{1}{i}[\hat{g}_{\mu \nu }(\mathbf{x},t),\hat{\pi}^{(3)\mu \nu }(\mathbf{%
x}^{\prime },t^{\prime })]_{t=t^{\prime }}=-(\frac{1}{2}\delta _{\mu }^{\nu
}\delta _{\nu }^{\mu }+\frac{1}{6}
\end{equation*}%
\begin{equation}
+2)\delta (\mathbf{x-x}^{\prime }\mathbf{)=}\Delta _{\mu \nu }^{(3)\mu \nu
}\delta (\mathbf{x-x}^{\prime }\mathbf{)}  \tag{7.5}
\end{equation}

Eqs.(7.3)-(7.5) mean that the cananical momenta of the general metric tensor
operators $\hat{g}_{\mu \nu }(\mathbf{x},t)$\ are the total momenta $\hat{\pi%
}^{(3)\mu \nu }(\mathbf{x}^{\prime },t^{\prime })$, the different
coefficients $-(\frac{1}{2}\delta _{\mu }^{\nu }\delta _{\nu }^{\mu }+\frac{7%
}{6}\delta _{\mu }^{\mu }\delta _{\nu }^{\nu }+g^{\mu \nu }g_{\nu \mu })$\
show their different commutation relations.

\section{VIII. Quantum gravity of the Universe}

\bigskip We now give applications of the general theory of quantum gravity
to gravitational fields of the Universe, namely, give quantum gravity of the
Universe.

For the key FRW metrics in current cosmology \cite{Jerzy}%
\begin{equation}
ds^{2}=dt^{2}-R(t)[\frac{dr^{2}}{1-\kappa r^{2}}+r^{2}d\theta ^{2}+r^{2}\sin
^{2}\theta d\varphi ^{2}  \tag{8.1}
\end{equation}%
where $R(t)$ is the scale factor of the universe and $\kappa =1,0,-1,$ which
correspond to closed universe, flat universe and negative curvature universe
respectively. Using eqs.(7.3) and (8.1), we have\bigskip

\begin{equation*}
\frac{1}{i}[\hat{g}_{00}(\mathbf{x},t),\hat{\pi}^{T00}(\mathbf{x}^{\prime
},t^{\prime })]_{t=t^{\prime }}=-(\frac{1}{2}\delta _{0}^{0}\delta _{0}^{0}+%
\frac{7}{6}\delta _{0}^{0}\delta _{0}^{0}+
\end{equation*}%
\begin{equation}
g^{00}g_{00})\delta (\mathbf{x-x}^{\prime }\mathbf{)=-\frac{8}{3}\delta (%
\mathbf{x-x}^{\prime }\mathbf{)}=}\Delta _{00}^{T00}\delta (\mathbf{x-x}%
^{\prime }\mathbf{),}  \tag{8.2}
\end{equation}%
\begin{equation}
\frac{1}{i}[\hat{g}_{rr}(\mathbf{x},t),\hat{\pi}^{Trr}(\mathbf{x}^{\prime
},t^{\prime })]_{t=t^{\prime }}\mathbf{=-\frac{8}{3}\delta (\mathbf{x-x}%
^{\prime }\mathbf{)}=}\Delta _{rr}^{Trr}\delta (\mathbf{x-x}^{\prime }%
\mathbf{),}  \tag{8.3}
\end{equation}%
\begin{equation}
\frac{1}{i}[\hat{g}_{\theta \theta }(\mathbf{x},t),\hat{\pi}^{T\theta \theta
}(\mathbf{x}^{\prime },t^{\prime })]_{t=t^{\prime }}\mathbf{=-\frac{8}{3}%
\delta (\mathbf{x-x}^{\prime }\mathbf{)}=}\Delta _{\theta \theta }^{T\theta
\theta }\delta (\mathbf{x-x}^{\prime }\mathbf{),}  \tag{8.4}
\end{equation}%
\begin{equation}
\frac{1}{i}[\hat{g}_{\varphi \varphi }(\mathbf{x},t),\hat{\pi}^{T\varphi
\varphi }(\mathbf{x}^{\prime },t^{\prime })]_{t=t^{\prime }}\mathbf{=-\frac{8%
\mathbf{\delta (\mathbf{x-x}^{\prime }\mathbf{)}}}{3}=}\Delta _{\varphi
\varphi }^{T\varphi \varphi }\delta (\mathbf{x-x}^{\prime }\mathbf{),}
\tag{8.5}
\end{equation}%
where $T$ is the total in the canonical momentum fields. The other
commutation relations of gravitational tensor fields $\hat{g}_{\mu ^{\prime
}\nu ^{\prime }}(\mathbf{x},t)$ and thier canonical total momentum fields $%
\hat{\pi}^{T\mu \nu }(\mathbf{x}^{\prime },t^{\prime })$ are zero, which and
eqs.(8.2)-(8.5) just show the homogeneity and isotropy of the Universe.

$\Delta _{00}^{T00}=\Delta _{rr}^{Trr}=\Delta _{\theta \theta }^{T\theta
\theta }=\Delta _{\varphi \varphi }^{T\varphi \varphi }=-\mathbf{\frac{8}{3}}
$ just show that our universe is the homogeneity and isotropy at large scale.

\section{IX. Quantum gravity of general black hole}

We now present applications of the general theory of quantum gravity to
gravitational fields of general black hole, namely, give quantum gravity of
general black hole.

For the gravitatiional fields of the most general black hole, i.e., the
metrics of Kerr-Newman black hole \cite{Jerzy}%
\begin{equation*}
ds^{2}=(1-\frac{2mr-Q^{2}}{r^{2}+a^{2}\cos ^{2}\theta })dt^{2}-\frac{%
r^{2}+a^{2}\cos ^{2}\theta }{r^{2}+a^{2}-2mr+Q^{2}}dr^{2}
\end{equation*}%
\begin{equation*}
-(r^{2}+a^{2}\cos ^{2}\theta )d\theta ^{2}-[(r^{2}+a^{2})\sin ^{2}\theta +
\end{equation*}

\begin{equation}
\frac{(2mr-Q^{2})a^{2}\sin ^{4}\theta }{r^{2}+a^{2}\cos ^{2}\theta }%
]d\varphi ^{2}+\frac{(2mr-Q^{2})a\sin ^{2}\theta }{r^{2}+a^{2}\cos
^{2}\theta }(dtd\varphi +d\varphi dt),  \tag{9.1}
\end{equation}%
then, eq.(9.1) can be equivalently and simply expressed as\bigskip
\begin{equation*}
g_{\mu \nu }=\left(
\begin{array}{cccc}
g_{00} & 0 & 0 & g_{0\varphi } \\
0 & g_{rr} & 0 & 0 \\
0 & 0 & g_{\theta \theta } & 0 \\
g_{\varphi 0} & 0 & 0 & g_{\varphi \varphi }%
\end{array}%
\right) \longrightarrow
\end{equation*}%
\begin{equation}
g^{\mu \nu }=\left(
\begin{array}{cccc}
\frac{g_{_{\varphi \varphi }}}{g_{00}g_{\varphi \varphi }-g_{0\varphi }^{2}}
& 0 & 0 & \frac{-g_{\varphi 0}}{g_{00}g_{\varphi \varphi }-g_{0\varphi }^{2}}
\\
0 & \frac{1}{g_{rr}} & 0 & 0 \\
0 & 0 & \frac{1}{g_{\theta \theta }} & 0 \\
\frac{-g_{_{0\varphi }}}{g_{00}g_{\varphi \varphi }-g_{0\varphi }^{2}} & 0 &
0 & \frac{g_{00}}{g_{00}g_{\varphi \varphi }-g_{0\varphi }^{2}}%
\end{array}%
\right)  \tag{9.2}
\end{equation}

using eqs.(7.3), (9.1) and (9.2), we achieve\bigskip
\begin{equation*}
\frac{1}{i}[\hat{g}_{00}(\mathbf{x},t),\hat{\pi}^{T00}(\mathbf{x}^{\prime
},t^{\prime })]_{t=t^{\prime }}=-(\frac{1}{2}\delta _{0}^{0}\delta _{0}^{0}+%
\frac{7}{6}\delta _{0}^{0}\delta _{0}^{0}+g^{00}g_{00})\delta (\mathbf{x-}
\end{equation*}%
\begin{equation}
\mathbf{x}^{\prime }\mathbf{)=-(\frac{1}{2}+\frac{7}{6}+\frac{g_{_{\varphi
\varphi }}g_{00}}{g_{00}g_{\varphi \varphi }-g_{0\varphi }^{2}})\mathbf{%
\delta (\mathbf{x-x}^{\prime }\mathbf{)}=}\Delta _{00}^{T00}\delta (\mathbf{%
x-x}^{\prime }\mathbf{)}.}  \tag{9.3}
\end{equation}

\begin{equation}
\frac{1}{i}[\hat{g}_{rr}(\mathbf{x},t),\hat{\pi}^{Trr}(\mathbf{x}^{\prime
},t^{\prime })]_{t=t^{\prime }}\mathbf{=-\frac{8}{3}\delta (\mathbf{x-x}%
^{\prime }\mathbf{)}=}\Delta _{rr}^{Trr}\delta (\mathbf{x-x}^{\prime }%
\mathbf{).}  \tag{9.4}
\end{equation}%
\begin{equation}
\frac{1}{i}[\hat{g}_{\theta \theta }(\mathbf{x},t),\hat{\pi}^{T\theta \theta
}(\mathbf{x}^{\prime },t^{\prime })]_{t=t^{\prime }}\mathbf{=-\frac{8}{3}%
\delta (\mathbf{x-x}^{\prime }\mathbf{)}=}\Delta _{\theta \theta }^{T\theta
\theta }\delta (\mathbf{x-x}^{\prime }\mathbf{).}  \tag{9.5}
\end{equation}%
\begin{equation*}
\frac{1}{i}[\hat{g}_{\varphi \varphi }(\mathbf{x},t),\hat{\pi}^{T\varphi
\varphi }(\mathbf{x}^{\prime },t^{\prime })]_{t=t^{\prime }}\mathbf{=-(\frac{%
5}{3}+}
\end{equation*}%
\begin{equation}
\mathbf{\frac{g_{00}g_{\varphi \varphi }}{g_{00}g_{\varphi \varphi
}-g_{0\varphi }^{2}})\delta (\mathbf{x-x}^{\prime }\mathbf{)}=}\Delta
_{\varphi \varphi }^{T\varphi \varphi }\delta (\mathbf{x-x}^{\prime }\mathbf{%
).}  \tag{9.6}
\end{equation}

\begin{equation*}
\frac{1}{i}[\hat{g}_{0\varphi }(\mathbf{x},t),\hat{\pi}^{T0\varphi }(\mathbf{%
x}^{\prime },t^{\prime })]_{t=t^{\prime }}=-(\frac{1}{2}\delta _{0}^{\varphi
}\delta _{\varphi }^{0}+\frac{7}{6}\delta _{0}^{0}\delta _{\varphi
}^{\varphi }
\end{equation*}%
\begin{equation*}
+g^{0\varphi }g_{\varphi 0})\delta (\mathbf{x-x}^{\prime }\mathbf{)=-(\frac{7%
}{6}-\frac{g_{\varphi 0}^{2}}{g_{00}g_{\varphi \varphi }-g_{0\varphi }^{2}}}
\end{equation*}%
\begin{equation}
\mathbf{)\delta (\mathbf{x-x}^{\prime }\mathbf{)}=}\Delta _{0\varphi
}^{T0\varphi }\delta (\mathbf{x-x}^{\prime }\mathbf{)\mathbf{=}\Delta
_{\varphi 0}^{T\varphi 0}\delta (\mathbf{x-x}^{\prime }\mathbf{)}.}
\tag{9.7}
\end{equation}%
The other commutation relations of gravitational tensor fields $\hat{g}_{\mu
^{\prime }\nu ^{\prime }}(\mathbf{x},t)$ and their canonical total momentum
fields $\hat{\pi}^{T\mu \nu }(\mathbf{x}^{\prime },t^{\prime })$ are zero,
which and eqs.(9.4) and (9.5) just show the symmetric properties of
Kerr-Newman black hole about diagonal elements in the covariant metrics of
eq.(9.2), eqs.(9.3), (9.6) and (9.7) just further show the effects of the
asymmetric property produced from the nondiagonal element field $\hat{g}%
_{0\varphi }$ in the covariant metrics of eq.(9.2) for Kerr-Newman black
hole.

Especially, we find that except the spatial singularities, there are the
singularities from the interactions $g_{00}g_{\varphi \varphi }=g_{0\varphi
}^{2}$ of metric fields, which just shows the nonuniformity related to the
metrics of time coordinate and angular coordinate.

\section{X. Summary and Conclusion}

This paper gives both a general canonical quantum gravity theory and the
general canonical quantum gravity theories of the Universe and general black
hole, deduces general commutation relations of the general gravitational
field operators and their different styles of high order canonical momentum
operators for this general nonlinear system of the standard gravitational
Lagrangian.

This paper concretely show the general commutation relations between the
general gravitational field operators and their zeroth, first, second and
third styles of high order canonical mementum operators for this general
nonlinear system of the standard gravitatioal Lagrangian, and then have
finished all the four styles of cannonical quantization of the standard
gravity. Especially, the novel equations (6.4) and (6.5) are deduced for the
first time, which reflect the nonlinear structure properties of the
commutation relations for the standard gravitatioal Lagrangian, i.e., this
paper discovers $\Delta _{\mu \nu }^{(i)\alpha \beta }$ $(i=0,1,2,3)$ and
their relations reflecting symmetric propertis of the standard nonlinear
gravitational Lagrangian.

Since quantum gravitational field theory is the very important foundational
theory for studying different gravitational field theories. Finally, the
conclusions are completely consistent with the existing quantum
gravitational field theories because of the general property of this paper's
no depending on the concrete metric. The other processings for the
quantization are similar to the usual quantization processings, thus we
don't repeat here.

So far as is known to all, the ultimate form of quantum gravity is what, all
people don't know. This paper presents a general canonical quantum gravity
theory that does not depend on the concrete metric, which provides the
general canonical quantum gravity theory for benefiting to find the final
theory of quantum gravity. Especially, when people want to further do the
second quantizatin, they need to use the results of the cannonical
quantization to finish the second quantization that gives the ultimate form
of quantum gravity, these have been done in our following works due to the
length limit of this paper. And because the general canonical quantum
gravity theory without dependence of the concrete metric in this paper is
the most general, which meet the requirements of the ultimate quantum
gravity.

Therefore, this paper give a simpler, direct physical and easily
understandable general canonical quantum gravitational theory don't
depending on any concrete metric models.

Acknowledgments: The work is supported by the U.S. Department of Energy,
contract no. DE-AC02-05CH11231, NSF through grants PHY-08059, DOE through
grant DEFG02- 91ER40681 and National Natural Science Foundation of China
(No. 11875081).

\end{document}